\documentclass[a4paper,onecolumn,11pt]{article}
\pdfoutput=1
\usepackage[utf8]{inputenc}
\usepackage[english]{babel}
\usepackage[T1]{fontenc}
\usepackage{amsmath}
\usepackage{hyperref}

\usepackage{tikz}  
 
\usepackage{fullpage}
\usepackage{amsfonts,amssymb}
\usepackage{amsthm}
\usepackage{tikz}
\usetikzlibrary{calc}
\usetikzlibrary{arrows}
\usepackage{tikz-3dplot}
\usepackage{color}  
\usepackage{hyperref}
\usepackage{bm}
\usepackage{float} 
\usepackage{appendix} 
\usepackage{lscape}
\usepackage{cite}
\usepackage{mathrsfs}
\usepackage{appendix}
\usepackage{setspace}
\usepackage{mathtools}
\usepackage[round]{natbib}

\theoremstyle{definition}

\begin{document}

\title{ Disappearing Without a Trace: The Arrows of Time in Kent's Solution to the Lorentzian Quantum Reality Problem} 
	 
  \author{Emily Adlam  \thanks{The Rotman Institute of Philosophy, 1151 Richmond Street, London N6A5B7 \texttt{eadlam90@gmail.com} }}

\maketitle

\abstract{Most existing proposals to explain the temporal asymmetries we see around us are sited within an approach to physics based on time evolution,  and thus they typically put the asymmetry in at the beginning of time in the form of a special initial state. But there may be other possibilities for explaining temporal asymmetries if we don't presuppose the time evolution paradigm. In this article, we explore one such possibility, based on Kent's `final-measurement' interpretation of quantum mechanics. We argue that this approach potentially has the resources to explain the electromagnetic asymmetry, the thermodynamic asymmetry, the coarse-graining asymmetry, the fork asymmetry, the record asymmetry, and the cosmological asymmetry, and that the explanations it offers may potentially be better than explanations appealing to a special initial state. Our hope is that this example will encourage further exploration of novel approaches to temporal asymmetry outside of the time evolution paradigm. }

\tableofcontents
\newpage
 
\section{Introduction}
The lesson of nearly two hundred years' worth of attempts to derive the second law of thermodynamics from microphysics is, as Callender succinctly puts it, `\emph{no asymmetry in, no asymmetry out}' \citep{sep-time-thermo}:  there is no possibility of deriving an objective asymmetry from a time-symmetric theory \emph{and nothing else}, so if we want to obtain an asymmetry, we must add  asymmetry somewhere. And because physicists from Boltzmann's era up until the present day  have largely worked within the time-evolution paradigm, where   the universe is conceptualised as something like a computer which starts from an initial  state and evolves forwards in time \citep{Wharton}, it has seemed inevitable that   to explain the various qualitative asymmetries we observe around us we must put some asymmetry in at the beginning of time -  for the initial state of the universe  is  the only point where the time-evolution picture allows us  freedom to adjust anything. Hence the past hypothesis \citep{pittphilsci8894, EARMAN2006399,Albert2000-ALBTAC}: the thermodynamic gradient is traditionally explained by postulating an initial state which is special in some way, for example because it is a `very low entropy' state. 

But there is something suspect about this reasoning, because the  time-evolution paradigm is plausibly a  consequence of the thermodynamic gradient. After all, why is it that we are inclined to model the universe in terms of laws which predict the future from facts about the present? Surely this is because we ourselves experience time as moving from the present to the future, so laws of this form are   relevant to our practical interests. Moreover, the fact that we have such experiences is surely to some degree a consequence of the fact that   we  form memories in a temporal direction aligned with the `record asymmetry', i.e. the fact that records typically form from the past toward the future; and it has often been suggested that the record asymmetry and the thermodynamic gradient are in some way related  \citep{HAWKING1993559}. So it would seem that we should not too blithely presuppose  the time-evolution picture in our attempts to explain various qualitative asymmetries, for fear of begging the question.

Indeed, as discussed  in  \citep{adlam2021laws,chen2021governing,Wharton}, many aspects of modern physics no longer seem like a good fit for the time evolution paradigm - and once we move away from this paradigm we no longer have an excuse for refusing to give an explanation for the `special' initial state. Typically, such refusals are grounded on the idea that explanations necessarily proceed by showing how a condition is produced by evolution from earlier conditions, so there can be no possible explanation for the initial state of the whole universe: \emph{`the whole enterprise of explaining global boundary conditions is suspect, for precisely the reasons Hume and Kant taught us, namely, that we can't obtain causal or probabilistic explanations of why the boundary conditions are what they are'.} \citep{Callender1997-CALROH-3} But if we don't consider time evolution to be fundamental, then there is no reason why we must always explain some condition by appeal to earlier conditions, so the initial state is just as likely to be susceptible to explanation as anything else. As  \cite{timebomb} compellingly argues, if we had reason to think the final state of the universe were as special as the initial one, we would certainly consider that fact in need of explanation; so if we don't presuppose a time-evolution description it seems quite natural to seek an explanation for the initial state.  

Thus,  given that accounts of temporal asymmetry based on the past hypothesis arguably have unsatisfactory features in any case \citep{Price2004-PRIOTO-2,Earman2006-EARTPH-3}, it seems worthwhile to see if there are other possibilities outside the time-evolution paradigm. In this article we explore an approach of this kind based on a proposed interpretation of quantum mechanics due to Kent \citep{Kent, 2015KentL, 2017Kent}. Specifically, Kent's model proposes that the content of reality is determined by the result of a `measurement' on the quantum state at the end of time, and thus we argue that the approach entails that the observed history of the universe is necessarily selected from a set of histories which are  strongly decoherent. We will argue that because of the special properties of these sets of histories, the final measurement is very likely to select a history exhibiting a number of the temporal asymmetries observed in our actual universe, and that moreover these temporal asymmetries are very likely to  be aligned in exactly the way we experience them.

We will ultimately suggest that this approach has a number of advantages over the past hypothesis, because in this picture the temporal asymmetries are derived from inherent features of the underlying model rather than simply attributed to an initial state which has been postulated in an ad hoc way. That said, we are not aiming to make the case that Kent's approach is definitely the right answer to the conundrum of temporal asymmetry - rather we present it as a  suggestive example which demonstrates clearly that  outside of the time-evolution paradigm there may be many unexplored possibilities for explaining temporal asymmetries, and we hope this example might encourage  further exploration of such  approaches.

\section{Kent's final-measurement interpretation}  

Kent's proposed interpretation of quantum mechanics is based on a simple idea. There is no collapse of the wavefunction; we just allow the wavefunction to undergo its standard unitary evolution over the whole course of history, and then, at the end of time, we imagine a single measurement being performed on the final state to determine the actual content of reality:  in Kent's words, `\emph{an event occurs if and only if it leaves effective records in the final time ... measurement}' \citep{Kent}. In heuristic terms, we might say that the final-measurement interpretation is bad news  for people engaged in nefarious schemes who hope to avoid leaving any evidence of their deeds - for it implies that events which leave no traces do not happen at all!

Kent has suggested several versions of this model.    \citep{2015KentL}  proposes   a measurement of mass-density across the final time hyperplane, with the actual contents of reality (the `beables')  given by two mass-density distributions  over spacetime for two classes of particle, such that the  value of the mass-density for one particle class at a given point $x$  is equal to the expectation value of the  mass-density operator for that class at $x$, conditional on the outcome of the final measurement on the other particle class.  Similarly,   \citep{  2017Kent} proposes measurements on the positions of photons which have been reflected off matter at various points, and then  defines the beables at $x$ by the expectation value of some operators (e.g. the stress-energy tensor components) at  $x$ conditional  on the detections of photos outside the future lightcone of $x$.  Note that   the only probabilistic component in Kent's approach is the final measurement: once an outcome has been selected, the records stored in the selected outcome determine the rest of the course of history uniquely, and anything it does not record simply doesn't occur. 

In this paper we will not distinguish between these models; we will simply understand Kent's approach to be defined by the following postulates (for concreteness we assume the beables are given by stress-energy tensor components, but using a mass-density or charge current or some other similar operator would not change any of our conclusions):

\begin{enumerate} 

\item We specify an initial state $| \psi \rangle_i$ and then the final state $| \psi \rangle_f$  is given by $U  | \psi \rangle_i$ where $U$ is the operator $e^{-iHt_f/\hbar}$ representing time evolution from $t = 0$ to $t = t_f$

\item A final measurement is performed, projecting on a set of outcomes $\{ O_x \}$, such that the probability for outcome $O_x$ is given by $  \langle \psi|_f O_x | \psi \rangle_f$ (the choice of the set $\{ O_x \}$ will depend on the specific model as described above)

\item Reality is composed of a  stress-energy tensor defined over a fixed spacetime background, with the value of the stress-energy tensor component $T_{uv}$ at the spacetime point $(x, t)$ given by $\langle T_{uv}(x,t) \rangle$, the expectation value of the stress-energy tensor component at the point $(x,t)$ conditional on the fixed initial state $\rho_i$ and the outcome $O_x$ of the final measurement (where the  way in which the model is conditioned on the outcome $O_x$ varies according to the specific model as described above)

\end{enumerate}

 We emphasize that the final measurement need not be performed by a literal observer external to the universe - the term `measurement' is simply a way of describing a process in which some particular course of history is selected and actualised. Indeed it need not even be the case that this `measurement' literally occurs at the end of time - if there is no end of time we can simply take the limit of the relevant probability distribution as $t$ goes to infinity, as detailed in   \citep{Kent}. Nor do we necessarily need to think of the evolution of the wavefunction as literally taking place with respect to some kind of pseudo-time parameter, for the `evolution’ is deterministic and thus specifying the state at any one time is equivalent to specifying it at any other time; so rather than postulating a literal temporal evolution of the wavefunction  one can simply specify the initial state and then use that to directly calculate the final probability, by choosing a  basis $\{U^{-1} O_x U \}$ for the final measurement which incorporates an appropriate unitary transformation encoding the effect of the `evolution'.

 As discussed by Kent in  \citep{Kent,2015KentL,2017Kent}, what makes this interpretation feasible is the existence of decoherence: as a quantum mechanical system interacts with its environment, forming correlations which rapidly disperse through the environment, interference terms are suppressed and the system tends towards a quantum state which has the form of a probabilistic mixture of classical states.  The rate of decoherence is roughly determined by the size of the system, so macroscopic systems typically undergo decoherence very quickly. Thus suppose the quantum state encodes several different possibilities for some classical property of a macroscopic system at  time $t$. Due to decoherence the state will quickly evolve towards a probabilistic mixture of these possibilities; and because the environmental correlations suppressing interference are widely dispersed so their effects cannot easily be undone, this probabilistic decomposition will typically be preserved under subsequent time evolution, so the final measurement can be expected to select exactly one possibility for the property at time $t$, thus defining beables which represent the system as having a single definite property at time $t$. This is the explanation offered by the final-measurement approach for the fact that we  perceive ourselves to be living in a universe full of macroscopic objects with  definite properties. 

On the other hand, in  laboratory experiments quantum-mechanical systems are kept isolated to prevent decoherence, so they are typically subject to subsequent interference effects. For example,   in the the double-slit experiment \citep{doubleslit} one branch of the wavefunction represents the particle passing through the left slit and another represents it passing it through the right slit,   but these branches interfere at the end of the experiment, so the subsequent state does not tend towards a probabilistic mixture of the two possible paths; this means  the final measurement won't determine which slit the particle went through,   so the beables   won't represent the particle as definitely going through either slit.  This is explanation offered by the final-measurement approach for the fact that  we are   able to see superposition and interference effects in sufficiently  isolated microscopic systems. 

  \subsection{Histories \label{histories}} 
  
In Kent's approach the role of the final measurement is to select and actualise some possible history of the universe -  the universe at the end of time is something like an enormous `register' whose quantum state has the form of a probabilistic mixture of different possible records of the whole of history, so measuring the register amounts to selecting exactly one  record of the whole of history, and the course of history that is actualised is the one specified by the selected record.

We will characterise the   kinds of histories that can be actualised by this mechanism using some terminology from  the consistent histories formalism \citep{gellmann1995strong,articlegriffiths}. In this formalism we define a `history' as a chain of time-indexed projection operators $E_{t_1}^i  , E_{t_2}^j    ...  s E_{tn}^k$ such that each projection operator $E_{t_i}^j$ is understood to represent an event taking place in the history at time $t$. To assign probabilities to these histories, we  specify an initial state $\rho$  and the probability for the history $E_{t_1}^i,   E_{t_2}^j   ...  E_{tn}^k$ is given by $ Tr( E_{t_n}^k ...  E_{t_2}^j   E_{t1}^i \rho  E_{t1}^i    E_{t_2}^j   ...E_{t_n}^k )$. A set of histories is `consistent' when this algorithm results  in a set of probabilities  such that the probabilities for mutually exclusive events can  be added up in the usual classical way. This will be the case provided the set satisfies the weak decoherence condition: the interference term $ Tr( E_{t_n}^k ...  E_{t_2}^j   E_{t1}^i \rho  E_{t1}^m    E_{t_2}^n   ...E_{t_n}^o )$ vanishes for any two distinct histories $E_{t_1}^i,   E_{t_2}^j   ...  E_{tn}^k$ and  $E_{t_1}^m,   E_{t_2}^n   ...  E_{tn}^o$ in the set \citep{https://doi.org/10.48550/arxiv.quant-ph/0701225}.

A key issue for consistent histories approaches  is that extending a history into the future can result in the loss of decoherence between past alternatives, so the past is not necessarily stable    \citep{DowkerKent}. One way to solve this problem is to insist on sets of histories which are not only consistent but also `strongly decoherent', a  condition defined by \cite{gellmann1995strong} to single out those sets which have  `\emph{records that guarantee the permanence of the past'}: at  each step in the chain of projections, the next projection is chosen from the set of projections that commute with generalized records of history up to that moment, so the projection leaves the generalized records undisturbed. `Generalized records' are   sets of orthogonal projection operators which distinguish between the different possible histories in a consistent set - i.e. they define measurements whose outcome would single out a unique   history from the consistent set. When generalized records exist at some time, the probability distribution over the possible outcomes of the measurement corresponding to the corresponding set of orthogonal projection operators at that time  is the same as the probability distribution over histories defined by the consistent histories formalism at that time. Generalized records don't have to be accessible to human observers -   Gell-Man and Hartle suggest that  \emph{`the physical picture is that, at every branching of the coarse-grained histories of the universe, each of the exhaustive and mutually exclusive possibilities is correlated with a different state of something like a photon or neutrino going off to infinity'.}  The idea is that when two histories diverge at time $t$, if some particle becomes correlated with events at $t$ such that the different events occurring at $t$ in these two histories correspond to orthogonal states of the particle,  we can then determine which of the histories we are in by measuring that particle - so   if the particle never interacts again, we will be able to distinguish between these histories at any time after $t$. Particles going off to infinity are not necessarily the  only possible way of arriving at a  strongly decoherent set of histories, but we will focus on this mechanism since it leads to a relatively intuitive picture.


Now, the simple version of the consistent histories formalism we have described here is not Lorentz covariant, while Kent's model is  Lorentz covariant, so the histories selected by the final measurement necessarily  have a more complex structure than the time-indexed chains of events appearing in the simple consistent histories formalism - in particular, Kent's model works with a continuous distribution over spacetime, rather than a sequence of discrete time-indexed events. However, the definitions we have given have natural analogues in  Kent's context, so we can use the consistent histories terminology to motivate some claims about the kinds of histories from which Kent's final measurement must select (though we emphasize that as a result of the structural differences the two formalisms cannot be exactly mapped into one another, so the use of the consistent histories terminology in Kent's framework has the status of an analogy rather than a precise mathematical identity).

First, postulate three tells us that Kent's model involves a selection from a set of `histories' defined by a distribution of stress-energy tensor components over spacetime, so the histories can each be thought of as corresponding to a set of projection operators associated with values for stress-energy tensor components at various spacetime points: that is, each outcome $O_x$ of the final measurement can   be mapped to a history $H_x$ such that the expectation value for each operator in $H_x$, conditional on $O_x$ and the fixed initial state, is approximately equal to $1$. Then postulate two defines a measurement  on the final state corresponding to a set of  orthogonal projection operators which distinguish between these possible histories, which means that it is  a measurement on a set of generalized records for these histories. Since the probability distribution over a measurement on a set of generalized records is, in the ordinary consistent histories case, the same as the probability distribution over the corresponding histories, we can think of Kent's model as something like a Lorentz-covariant formulation of the consistent-histories approach  which also prescribes a specific way of selecting one consistent set out of all of the possible consistent sets. In particular, one important feature of Kent's specification  is that it by default chooses a set of histories which preserves the permanence of the past, since an event can only feature in one of the histories from which the final measurement selects if it is recorded in the generalized records on which the final measurement projects. This means the set of histories from which the final measurement selects must be a strongly decoherent set, or rather it must be some generalisation of a strongly decoherent set which is appropriate to Kent's continuous and Lorentz-covariant context. Specifically, it  must exhibit this generalized strong decoherence with respect to the temporal direction defined by the stipulation that the `final measurement' occurs at the end of time, and therefore in the context of Kent's model the temporal asymmetry in the definition of strong decoherence is not an independent temporal asymmetry, but is simply a consequence of the asymmetry inherent in the final-measurement picture.

Following Gell-Man and Hartle, we will  assume  that the generalized kind of strong decoherence exhibited by the set of histories associated with the final measurement is such that  for every history $H_x$ in the set and  every event $E_i$ in that history, there is at least one system carrying a record of $E_i$ which does not interact again after $E_i$. This assumption is consistent with the approach taken by  \cite{2017Kent}, where the final measurement takes the form of a position measurement  on a set of photons: in this model each photon can be understood as carrying a partial record of the  system it last interacted with, with the complete information stored in the correlations between the final positions of all of the photons, so we do indeed have a set of record-carrying systems which never interact again after the record-forming interaction.

\section{The Arrows of Time  \label{arrows}}

Kent's approach looks like a promising place to seek new answers to old questions about the arrow of time, because it  combines an underlying time-symmetry (the unitary evolution of the quantum state) with an objective asymmetry (the actual content of reality is determined by the final measurement, which occurs at one   end of time). Thus it points to a way of explaining why our reality appears so asymmetric despite the time-symmetry of the underlying dynamical equations.

As observed by   \cite{Earman1974-EARAAT}, it is important in these discussions to be precise about exactly what problem we aim to solve. For there is not just  one  arrow of time - the universe exhibits a variety of temporal asymmetries, which must be treated separately unless we can demonstrate that they are in fact aspects of the same asymmetry.  In this article we will not seek to explain why these `arrows of time' point in one direction rather than another - we don't believe that there is anything outside the universe relative to which the direction could be measured, so we don't consider this question to be meaningful. We also will not seek to explain why various different arrows of time all point in `the same direction', because the choice to describe a temporal asymmetry as `pointing' in one direction rather than another is a conventional one. Thus our aim is, first, to explain why these temporal asymmetries exist at all, and second, to explain why they are all aligned in the specific way that they are in our actual world - for example, why entropy increases and electromagnetic waves spread out in the same temporal direction.

As reinforced by Callender's mantra, in the context of quantum mechanics these asymmetries can't be explained by the time-symmetric unitary evolution, so we must add some asymmetry.  The usual way of doing this is to choose a `special' initial state - for example, \cite{https://doi.org/10.48550/arxiv.2211.03973} suggests an initial state given by a projection on the `Past-Hypothesis subspace'. But alternatively, since the unitary evolution produces a large number of branching histories and our observed reality corresponds to only one history, we may use a generic initial state and then select out the history corresponding to our  actual observed reality in such a way that it is  likely to display some characteristic asymmetry. \cite{https://doi.org/10.48550/arxiv.quant-ph/0701225} advocates an approach of this kind using the consistent histories formalism, but this proposal is not a complete explanation of temporal asymmetry, since it is still necessary to adopt some specific interpretation of quantum mechanics in order to say how the various histories that can be extracted from the wavefunction relate to our actual experiences.  On the other hand, Kent's final-measurement interpretation allows us to implement this approach in a natural way, since it explicitly specifies a set of histories from which our observed reality must be selected.
 
 Thus we will henceforth assume that in our actual universe the initial quantum state (pure or mixed) encodes no particular temporal asymmetry, so the overall quantum description is roughly time-symmetric, but nonetheless many different types of histories can be extracted from it: some  in which there is no temporal asymmetry at all, some exhibiting an `arrow of time' oriented in one direction, some exhibiting  the same `arrow of time' oriented in the other direction. We will argue that a number of temporal asymmetries can be derived from the assumption that the final measurement selects from a strongly decoherent set of histories, where we take it that such a set is likely to consist of histories in which every event is associated with a record system that goes off to infinity and never interacts again. 
 
 Of course, if the histories in question are very short  this requirement will not necessarily lead to any qualitative asymmetry - e.g. for  a set of histories each consisting of just a single event, it is automatically the case that there is a record of that event stored in the state of a system which never interacts again, so the histories don't need any asymmetry to achieve this. But if we are considering long complex histories the requirement is quite non-trivial, because in this case the systems carrying records of a given event will subsequently have many opportunities to interact.  And thus in order to ensure that all events in the history produce at least one record which doesn't interact again, we will typically require that each event is recorded in a large number of different systems, so it is likely that at least one record will survive. Thus if we assume that our universe has a long  history, then the requirement of strong decoherence strongly favours histories  in which events typically produce many records oriented  towards the future.    We will aim to show that  several of the `arrows of time' observed in our world can be explained in this way, and that moreover we predict that they should all be aligned in exactly the way they are aligned in the actual world.

 \subsection{The Electromagnetic Asymmetry}
 
 The `electromagnetic asymmetry' refers to the fact that we frequently observe emitters producing electromagnetic waves which spread out into the environment, whilst we seldom seem to observe a collection of electromagnetic waves precisely converging on a single absorber \citep{Price1991-PRITAO-7}. Yet the latter process is the time-reverse of the former, and moreover the equations of electromagnetism are time reversal invariant, so converging to absorption is  physically possible, and if we look only at the equations of electromagnetism it would seem that both kinds of processes should be equally likely. 
 
 We can explain this asymmetry within the final-measurement interpretation.  An electromagnetic wave which spreads out and is absorbed by a large number of different absorbers will leave many records, since each one of the absorbers contains a record of the wave it has absorbed (for example, it may undergo a change in energy levels as a result of the absorption), and thus there is a good likelihood that at least one of the record-carriers will subsequently never interact again. Alternatively, if the wave happens to be in a region where there are no absorbers, then the wave itself will simply spread out and continue traveling without further interactions until it arrives at the end of time, providing a record of its own history. Thus for any  strongly decoherent set of histories where the histories contain events pertaining to electromagnetic waves, it is likely that the histories will feature a lot of emission events where electromagnetic waves spread out into the environment.  Conversely, if a collection of waves converges on and are absorbed by a single absorber, the only record of their passage is stored in the state of that single absorber, which makes the record very vulnerable: there is a significant chance that this one absorber will interact with something else and then all records of the waves will be destroyed. So  a set of histories consisting of histories  which  contain a lot  of instances of waves converging on a single absorber is not likely to be strongly decoherent\footnote{This way of thinking about the electromagnetic asymmetry does have the interesting consequence that, although we shouldn't expect to see many instances of waves converging on an absorber, nonetheless if we look in the right places we should sometimes be able to catch waves converging towards an absorber, since the act of looking creates a macroscopic record which could persist until the end of time. However, in practice this would be quite hard to do since we have no way to know the right time to look for the waves or the directions from which they are converging, so it's unclear how one could design an experiment to look for such an effect.}.
 
Thus the final-measurement interpretation gives us a picture where we do indeed have time-symmetry with respect to electromagnetic processes at the level of the unitarily evolving wavefunction - if we could see the full wavefunction,  we would find that it represents  waves converging  on single absorbers just as often as waves emerging from single emitters, as expected from the time-symmetry of the equations. But  that symmetry is broken by the final measurement, because the measurement has to select from a strongly decoherent set, so it is likely to select a history  containing many more emission events than absorption events.

 \subsection{The Thermodynamic Asymmetry \label{thermodynamics}} 
 
 The `thermodynamic asymmetry' is familiar to us from everyday life - plates tend to break but very seldom do pieces of broken china fly back together to form a plate, gases tend to diffuse to fill a room but very seldom do they shrink into the corner, and so on.  These phenomena are unified by the second law of thermodynamics, which in its most well-known form states that the total entropy of the universe must always increase \citep{sep-time-thermo}: entropy is, roughly speaking, a measure of disorder, so glasses breaking, gases diffusing and so on are typically regarded as entropy-increasing processes.  
 
 Why should it be the case that entropy so reliably increases if the underlying laws are time-symmetric? The orthodox explanation of this asymmetry begins by postulating that the initial macrostate of the universe was a special `low entropy' macrostate  and then argues that since low entropy macrostates occupy a significantly smaller region of state space than large entropy macrostates, all else being equal we would expect the universe to evolve towards a higher entropy macrostate. This assumption about the special initial macrostate is known as the `past hypothesis'. \citep{Albert2000-ALBTAC}  But concerns have been raised about the past hypothesis approach - including the worry that it seems like a piece of unreasonable good luck that the universe should have happened to have this very special low entropy initial macrostate \citep{Price2004-PRIOTO-2}, and the worry that the low entropy state is too far in the past to explain irreversible behaviour in the present \citep{EARMAN2006399}. We will  discuss the past hypothesis in more detail in section \ref{ph}, but for now let us consider if the final-measurement approach can offer a different account.
 
 It is sometimes argued that the creation of records is necessarily associated with entropy increase: for example, \cite{HAWKING1993559} writes, \emph{`when a computer records something in memory, the total entropy increases. Thus computers remember things in the direction of time in which entropy increases. In a universe in which entropy is decreasing in time, computer memories will work backward'.}  If this is correct, then we have our answer. For if it is true that entropy increases every time a record is formed, then a history in which events are usually followed by the formation of many records will be a history in which entropy reliably increases towards the future; and we know that the final measurement requires the formation of many records oriented towards the future, so it seems natural to think the final measurement is likely to select a history in which entropy reliably increases towards the future.

But is it true that the formation of records is inevitably associated with an entropy increase? Arguments to this effect largely invoke Landauer's Principle: \emph{`To erase a bit of information in an environment at temperature $T$ requires dissipation of energy greater than $ kT \ln(2)$'.} \citep{pittphilsci2689} It is then argued that recording something in memory requires the erasure of information, since the process will necessarily overwrite whatever was previously written in the register used for the record, so the creation of records involves an increase in entropy. However, this whole argument is quite suspect, because it remains  unclear that   Landauer's Principle is universally valid. Now, for our purposes we probably don't need  Landauer's Principle to be universally valid - to explain an overall thermodynamic gradient it would be enough to show that most realistic record-forming processes   involve erasure and that  Landauer's Principle applies in most of these processes, and this does appear to be a true generalisation in our actual world. But in order for this generalisation to play the required role in a derivation of the thermodynamic gradient, it would need to be the case that entropy is increased in  these processes purely because they are record-forming processes, not because we live in a universe in which entropy increases along with the direction of record-formation (that would be begging the question!). And  in fact, it turns out that arguments for Landauer's principle typically presuppose that we live in an entropy-increasing universe \citep{pittphilsci2689}  - as shown by   \cite{mar2009}, in an entropy-decreasing universe the usual kind of derivation leads to the conclusion that  `\emph{to erase a bit of information in an environment at temperature $T$ requires dissipation of energy less than - $ kT \ln(2)$}' -  so if our purpose is to explain the existence of an entropy gradient, we are not entitled to simply presuppose Landauer's Principle in the way that the argument suggested above requires.

So in fact, we suggest that the explanation of the thermodynamic arrow within the final-measurement interpretation is based not on the record-forming process itself, but rather on the subsequent dispersal of those records into the environment. For the characteristic feature of decoherence is not merely that events leave records in the environment, but that those records rapidly `spread' so that it would be very difficult to access and control all of the records in order to undo the effects of the event, which greatly increases the probability that there will be at least  one record-carrying system which never interacts again. This feature is crucial for strong decoherence, because  if all of the records of an event remain confined to a single system in its local environment, then those records will probably not survive  until the end of time: it is far too likely that something will happen to damage the single system containing all the records. Thus since the final measurement must select from a strongly decoherent set of histories, it is likely to produce a history   in which the dispersal of microscopic records into the environment is ubiquitous. 

Let us now make use of a characterisation of entropy increase advocated by \cite{mar2009}: `\emph{In an entropy increasing universe ... the microscopic correlations that develop due to  $\phi^{(t)}$ play no role in the future evolution of the system}'. Here, $\phi^{(t)}$ refers to the microscopic time evolution prescribed by the relevant dynamical laws. Now, in our application we are interested not in entropy increasing universes but in entropy increasing histories: according to Maroney's characterisation, in such histories, microscopic correlations by and large \emph{`play no role in the future evolution',} within the history, which means that these correlations quickly become inaccessible. In Maroney's words, `\emph{The coarse graining over the microcorrelations, that corresponds to entropy increase, is associated with the inaccessibility of these microcorrelations. If the microscopic correlations were still accessible  ... no entropy increase could be said to have occurred.}'  Note that whenever the records of an event take the form of correlations which are `inaccessible', this means that the microscopic systems carrying the records cannot be disturbed, destroyed or contained by macroscopic interventions, i.e. they are robust against macroscopic perturbations, which means there is a good chance that some of these systems will eventually escape their immediate environment and `go off to infinity', never interacting again. This is exactly what we need for strong decoherence, so we may conclude that a strongly decoherent set of histories is likely to contain mostly histories which are reliably entropy-increasing.  

Conversely,   \cite{mar2009} characterises entropy decreasing universes as those in which `\emph{microscopic correlations disappear over the course of the interaction}'. Thus in histories which are entropy-decreasing, records of events  stored in correlations typically vanish altogether over time, meaning that a set composed of histories which are entropy-decreasing   will generally fail to be strongly decoherent. So the final measurement approach predicts that the history selected by the final measurement will most likely contain many entropy-increasing processes and few entropy-decreasing processes; which is to say, it tells us that we should expect to find ourselves in a history which has an overall   entropy gradient and which therefore exhibits typical thermodynamical behaviour.

One intriguing   consequence of this argument is that it  suggests the rate of entropy increase may not be constant. The reason records must be dispersed rapidly into the environment in the final measurement approach is we need to have at least one record which never interacts again; so if an event occurs very shortly before the end of time, there is no need for its records   to disperse into the environment because  there won't be many opportunities for the records to be damaged.  Conversely, if an event occurs very early in the history of the universe, then its records must be very robust indeed if they are to survive until the end of time. And in fact,  according to our current understanding of the early universe, it seems that  many events in the early history of the universe do indeed produce very robust records:  small fluctuations in the density of the early universe are hypothesized to have given rise to very large structures like galaxies and galaxy clusters, which can be regarded as `records' of them \citep{cimatti2019introduction}. Thus the records of these fluctuations have indeed become inaccessible via dispersal into the environment, though instead of simply becoming invisible as microscopic correlations typically do, they have instead formed extremely robust large-scale structures which are so spread out that there is no feasible way that the records they encode could ever be erased. Of course some care must be taken here to avoid bias in this analysis - obviously the records that survive from the early universe until the present day will be the more robust ones, so there is a danger that we will get the impression that records created earlier in time are on average more robust than records created in the present simply because we don't have any way to know about all the less robust records created in earlier in time that did not survive. However,  according to our current understanding of cosmology it does seem to be the case that there is no mechanism comparable to inflation which could encode present-day density fluctuations in large-scale features of the universe like galaxies, so there is  at least a case to be made that records are becoming less robust as time goes on, exactly as the final-measurement interpretation  seems to predict. It would be interesting to see if this argument could be made more quantitative.

\subsection{The Coarse-Graining Asymmetry\label{coarse}} 

The `coarse-graining asymmetry' is one particular manifestation of the thermodynamic asymmetry: as urged by  \cite{pittphilsci8894},  in most realistic situations  when we want to make predictions, we don't use the precise microscopic dynamics but rather employ a coarse-grained dynamics which allows us to describe macroscopic evolution directly. There are various different coarse-graining techniques, but they share a basic structure as described by \cite{pittphilsci8894}:  `\emph{Firstly, we identify a set of macroproperties ... in whose evolution we are interested. Secondly, we define a map C — the coarse-graining map — which projects from the distribution space onto some subset $S_C$ of the distributions ...  We then define the forward dynamics induced by $C$  ... as follows: take any distribution, coarse-grain it, time-evolve it forward (using the microdynamics) by some small time interval $\delta t$, coarse-grain it again, time-evolve it for another $\delta t$ and so on.}'  The crucial point is that the resulting coarse-grained dynamics is generally irreversible, so if the dynamics make correct predictions when applied in the forwards time direction, they will not generally make correct retrodictions when applied in the backwards time direction. And in fact these coarse-grained approaches are known to be very successful at making accurate predictions when applied in the forward temporal direction, which gives us another asymmetry to explain: why does coarse-graining work in the forward direction but not generally in the backward direction? 

The final-measurement approach is relevant to the coarse-graining arrow because  coarse-graining is often used to calculate the effect of environment-induced decoherence: this is done by `\emph{alternating unitary (and entangling) interactions of system and environment with a coarse-graining defined by replacing the entangled state of system and environment with the product of their reduced states.}' \citep{pittphilsci8894} Now, we saw in the previous section that if we wish to have a good chance that at least one of the records of an event never interacts again, the records of that event must be robust against macroscopic perturbations, i.e. they must spread out into the environment in such a way that they can't easily be disturbed, destroyed or contained by macroscopic effects.  This means that in general the correlations storing records of past events must play no role in the macroscopic evolution of the system, for if they did have an effect on the macroscopic evolution of the system, then macroscopic events  would also have an effect on them so could be used as `handles' to disturb, destroy or contain them. And Maroney notes that if \emph{`the microscopic correlations that develop due to  $\phi^{(t)}$ play no role in the future evolution of the system'} it follows that \emph{`we may make the coarse grained replacement $\phi^{(t)}(\delta_0) \rightarrow  \cup_n\vec{\delta}_n  \otimes \delta E$ for all future evolution of the system',} because the effect of the coarse-graining technique is precisely to discard microscopic correlations.  That is to say,  whenever an event is followed by evolution which encodes records of it in the environment in such a way that it's very likely some of the records never interact again, it will typically  be the case that we can successfully describe that evolution at a macroscopic level using a coarse-graining procedure which discards microscopic correlations. Thus the final-measurement account  also  explains why coarse-graining works:  the final measurement must select from a strongly decoherent set of histories and such histories must contain robust records which are likely to persist until the end of time, which also means that the coarse-graining approximation will mostly be valid.

\subsection{The Fork Asymmetry \label{fork}}

\begin{figure}
	\centering
	\begin{tikzpicture}[scale=0.7]
	
	\coordinate (a) at (0,0);
	\coordinate (c) at (0,2);
	\coordinate (d) at (2,2);
	\coordinate (b) at (-2,2);

	\draw[black, arrows={-triangle 90}] (a) -- (b);
	\draw[black, arrows={-triangle 90}] (a) -- (c);
	\draw[black, arrows={-triangle 90}] (a) -- (d);

	\node[below, black] at (a) {$A$};
\node[above, black] at (b) {$B$};
\node[above, black] at (c) {$C$};
\node[above, black] at (d) {$D$};

	\end{tikzpicture}	
	\caption{Future-directed fork}
	\label{figsig}
\end{figure}
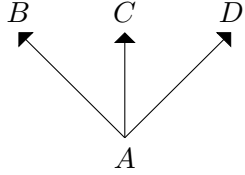

\begin{figure}
	\centering
	\begin{tikzpicture}[scale=0.7]
	
	\coordinate (a) at (0,0);
		\coordinate (a1) at (-0.3,0);
			\coordinate (a2) at (0.3,0);
	\coordinate (c) at (0,-2);
	\coordinate (d) at (2,-2);
	\coordinate (b) at (-2,-2);

	\draw[black, arrows={-triangle 90}] (b) -- (a1);
	\draw[black, arrows={-triangle 90}]  (c) -- (a);
	\draw[black, arrows={-triangle 90}]   (d) --(a2);

	\node[above, black] at (a) {$A$};
\node[below, black] at (b) {$B$};
\node[below, black] at (c) {$C$};
\node[below, black] at (d) {$D$};

	\end{tikzpicture}	
	\caption{Past-directed fork}
	\label{figsig2}
\end{figure}
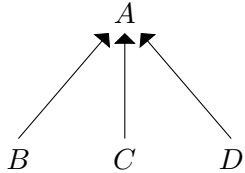

The `fork asymmetry' \citep{Lewis1979-LEWCDA,Horwich1987-HORAIT-2}, (sometimes also known as Reichenbach's principle \citep{sep-physics-Rpcc}), refers to the widespread existence of `forks', i.e. instances where two correlated events are statistically independent once we condition on a past event. For example, Bob's consumption of alcohol may be followed by enthusiastic dancing in the coming hours and then headaches the following morning. Dancing in the evening and headaches the next morning are   correlated, but when we condition on the consumption of alcohol they become statistically independent, because the dancing  is not the cause of the headaches or vice versa; rather they share a common cause in the consumption of alcohol. There appears to be an asymmetry in the way in which such forks feature in our universe - for \emph{`whenever two observable events A and B are correlated, there is often some earlier observable event C whose occurrence (and non-occurrence) renders A and B statistically independent, whereas there is never a later observable C that plays the same role'} \citep{pittphilsci19073}.  That is, in our universe correlated events usually have a common cause in the past, so our universe contains many `forks' as in the causal diagram shown in \ref{figsig}, but it's quite uncommon in our universe to see a pair of events `converging' on some  future event,  which would give a backwards fork as in figure \ref{figsig2}.

The fork asymmetry may at first seem to have an easy explanation within the final-measurement interpretation. For we have argued that the strongly decoherent set of histories from which the final measurement selects can be expected to contain histories in which events typically produce a large number of  records; and when an event $E$ is recorded in more than one physical system, we have something like a`fork', since if the records are accurate then they must be correlated but also statistically independent conditional on the event $E$. This indicates that histories featuring in a  strongly decoherent set  will generally contain many `forks' towards the future.  Moreover, there is a strong preference for forwards forks having a very large number of `prongs', because the more `prongs' the forks has, the more likely that at least one of the record-carrying systems will not interact again. 

However we must be careful here, for as emphasized by  \cite{10.2307/192839} and  \cite{10.1093/0195172159.001.0001}, the fork asymmetry pertains to forks composed of observable macroscopic events, whereas the generalized records defining a strongly decoherent set may  be stored in microscopic correlations, so they don't necessarily correspond to observable forwards forks. Still, it seems reasonable to expect that the ubiquity of microscopic forward forks   will have the consequence that we also see a significant number of macroscopic forward forks. For suppose we have a macroscopic event $E_M$ which is constituted by a collection of correlated and collocated microscopic events $\{ e_m\}$. Each such event $e_m$ must be recorded in the final state of the universe, so each must produce a large number of microscopic records of itself.   And since the events $\{ e_m\}$ are correlated and collocated, the systems carrying their records will also initially be correlated and collocated, so it is likely that as they spread into the environment they will undergo collective interactions in such a way as to produce further observable macroscopic events $F_M$ -  indeed, many such events, since  we saw in section \ref{thermodynamics} that the final measurement mechanism favours histories in which  microscopic records tend to disperse out into the environment, so the systems carrying records of the events $\{ e_m\}$ can be expected to be involved in a variety of different macroscopic events $\{ F_M \}$ in different regions of the environment.   These events  $\{ F_M \}$ will typically be correlated, since for two such events $F_M, F_M'$ there will in general be significant overlap in the set of microscopic events from $\{ e_m\}$ whose records are involved in the events $F_M$ and $F_M'$ respectively, but they will  be statistically independent conditional on the original event $E_M$, since  their correlations stem from their common origin in records of the events making up $E_M$; so $E_M$ together with the later events  $\{ F_M \}$ can be expected to form an observable forwards fork. Thus the final  measurement is indeed likely to give us a history containing many observable forwards forks.

What about the nonexistence of backwards forks? Well, suppose we did have a backward fork, as pictured in figure \ref{figsig2}. The final measurement approach tells us that in order for each of the observable events  $B, C, D$ to feature  in the actual course of history, they must all be part of a history belonging to the strongly decoherent set from which the final measurement selects,  so there must be records of them which persist until the end of time. So we need each of the microscopic events making up the macroscopic events $B, C, D$ to produce   a large number of records, and we have argued that this is likely to result in each of $B, C, D$ having its own observable forward fork.  Thus for every backwards fork where an observable event produces $n \geq 2$ observable `prong' events in its past, we will usually end up adding $n$  additional observable forward forks; so the histories belonging to a strongly decoherent set  can be expected to have many more observable forwards forks  than observable backwards forks, leading to a fork asymmetry.

 We should now address a criticism due to  \cite{arntzenius1990physics}, who noted that in a  deterministic theory, every event $C$ at time $t$ has a `determinant' at every other time - a set of physical conditions which are necessary and sufficient for the occurrence of $C$. So if it is the case that the occurrence of $C$ renders some future events $A$ and $B$ statistically independent, then at every other time $t$, including times later than the occurrence of $A$ and $B$, there exists a determinant of $C$, which must also render $A$ and $B$ statistically independent. So Arntzenius argues that there can be no fork asymmetry in a deterministic theory. 
 
The final-measurement approach is not deterministic, but Artnzenius' criticism still applies, because in the final-measurement picture  any event  featuring in the actual course of history must have generalized records at all later times, and since generalized records are by definition determinants of the events they record.   However,   the fork asymmetry pertains to observable events, and although we have seen that in the final measurement interpretation an observable event $E_M$ will often be followed by correlated observable events $\{F_M\}$ which can be regarded as records of  $E_M$, the events   $\{F_M\}$ will not typically be determinants of $E_M$, because they are only loosely-defined aggregates of some subset of the systems carrying generalized records of the microscopic events making up $E_M$, so they are unlikely to contain enough information to perfectly reconstruct $E_M$. Thus although $E_M$ would indeed have a determinant if we were to collect together all the generalized records of all of the microscopic events making up $E_M$, that collection will not usually comprise any well-defined single macroscopic event, so this does not correspond to an observable backwards fork.  So the final measurement does indeed predict a fork asymmetry at the level of observable macroscopic events, since  histories belonging to a strongly decoherent set are likely to  contain  more  observable forward forks than  observable backwards forks.

 \subsection{The Record Asymmetry \label{record}} 
 
The `record asymmetry' (sometimes also known as the `epistemic asymmetry') refers to the fact that we frequently have access to detailed and highly reliable records of the past, whereas we don't typically have such records of the future \citep{pittphilsci19073}. Of course, sometimes we can predict certain facts about the future from facts about the present, so one may regard those facts as in some sense providing `records' of the future; but it is clear that records of the past are in general much more detailed, accurate and widespread. This asymmetry is given particularly clear expression in the fact that we remember the past but not the future, for a memory is after all a kind of record. The record asymmetry is therefore closely tied to the memory asymmetry, and thereby to the psychological arrow of time, since clearly the fact that we experience time as moving from the past towards the future has at least something to do with the fact that we remember the past and not the future. 

Because records play such a prominent role in the final-measurement interpretation, it is straightforward to see where the record asymmetry comes from within this picture.  Kent's model tells us  that the final measurement must select a history from a strongly decoherent set of histories, so it predicts that we will find ourselves in a history in which events are generically followed by the formation of many different records of those events, and moreover those records should all form in the same temporal direction, i.e. towards the end of time where the final measurement occurs. Of course many of those records may be unobservable, but as we argued in section \ref{fork}, given any macroscopic event $E_M$, the aggregate of all the   forward-directed records of the microscopic events composing  $E_M$ can be expected to produce a  set of  macroscopic events in the future which can be regarded as records of $E_M$,  albeit records which are not determinants of $E_M$. So the final-measurement approach also predicts that we will find ourselves in a history in which   macroscopic events are  generically   followed by the formation of a significant number of observable records of those events, although the records will usually be  approximate and imperfect. 

One might think that we could  have records forming both towards the future and also towards the past. But  as in section \ref{fork}, the final measurement approach tells us that in order for the formation of a record toward the past to be a part of the course of history, there must be future-directed records of the formation of that backward-directed record which persist until the end of time, and indeed usually there will be many different future-directed records of it, so we can expect to get many more future-directed records than past-directed ones overall. Furthermore,  as argued  by  \cite{pittphilsci19073}, the fact that events typically form many different forwards-directed records is an important part of the reason why we are usually able to obtain reliable and detailed information about the past, as compared to the relatively sketchy information we can obtain about the future. So the histories belonging to a strongly decoherent set are likely to have the feature that at any given point in the history it is significantly easier to find out about the past than about the future, thus the final measurement  is likely to select a history exhibiting a record asymmetry. 

 \subsection{The Cosmological Asymmetry}

Kent's existing models for the final-measurement interpretation presuppose a Lorentzian background spacetime,  so all of the histories associated with possible outcomes of the final measurement live in the same spacetime. However, in a general-relativistic version of Kent's model it seems natural to suppose that the histories could  correspond to different spacetimes and cosmologies. For example, solutions to the Wheeler deWitt equation may include superpositions of different cosmologies, such as expanding and contracting universes - see ref  \citep{Craig_2010} - so if we suppose that a general relativistic version of Kent's model would involve performing a final measurement on a universal state $|| \Psi \rangle$ which is a solution to the Wheeler DeWitt equation, it follows that there will generally be several different cosmologies which could in principle be selected by the final measurement.  

Evidently a version of Kent's model capable of realising this idea would require a different kind of formalism - without a  fixed background spacetime, the beables could no longer simply be defined at fixed spacetime points as in Kent's existing models, so it's likely that these pointlike beables would have to be replaced with some other kinds of beables.   It is beyond the scope of this paper to construct such a model, which would in effect constitute a novel theory of quantum gravity. However, we will now make a few (highly speculative) comments about how the  cosmological asymmetry might be explained by a  general-relativistic version of Kent's model which allows choosing between different cosmologies, if such a thing could be constructed. 

Cosmology exhibits a number of notable asymmetries, but perhaps the most obvious is the fact that the universe is expanding - and, according to many current models, will continue to expand indefinitely \citep{possel2017expanding}.  There is definitely potential to explain this in a generalisation of Kent's model, because Kent's construction involves a measurement   at the end of time, and therefore it necessarily makes some claims about what the future looks like in cosmological terms. In particular, if it is possible to determine the whole course of history from the result of the final measurement, everything that happens now must be recorded in the state at the end of time, and therefore the final measurement cannot possibly select histories which involve  wholesale destruction of records before the end of time. Evidently this rules out cosmologies involving some kind of `big crunch', as collecting the material contents of the universe together into a small region will lead to a large number of interactions which will tend to destroy any records that might have been stored in the states of particles and photons.  In Kent's words, `\emph{We do not believe it is true in all cosmological models that quasi-classical systems leave effective records outside their future light cone. For example, models with a final ‘big crunch’ do not have this property'} \citep{2015KentL}. 

Indeed, it seems reasonable to say that not only must the final measurement select from histories which don't exhibit  big crunches,  in fact it must select from histories incorporating some kind of mechanism such that over time, more and more of the material contents of the universe cease to interact with the other parts. To see this, consider a simple toy example of a universe containing a finite number of   particles which are able to store records, such that every time one of these particles interacts with some system, it creates a partial record of the state of that system (with the full information obtainable from correlations from the records stored by various different particles) and any records it previously stored are erased. Let us divide these particles up into a set $N_I$ of particles which are still interacting with other particles and systems, and a set $N_N$ which have ceased to interact forever. As set out in section \ref{histories}, we are  assuming that in a strongly decoherent set of histories, all the histories have the property that for every event in the history there is at least one partial record of that event stored in a system which never interacts again. So in these histories, whenever an event occurs, at least one particle   switches from the set $N_I$ to the set $N_N$, which means that  the ratio $\frac{N_N}{N_N + N_I}$ must be constantly increasing across all of time. This suggests that the final measurement is likely to be choosing from a set of histories in which   $\frac{N_N}{N_N + N_I}$ is constantly increasing. 

Moreover, how is it that a particle can become non-interacting? Arguably the only really reliable way to ensure that a particle does not interact after some point in time is to make sure the particle becomes widely separated from other particles in order that it doesn't encounter anything to interact with. And thus one natural way to create a history in which $\frac{N_N}{N_N + N_I}$ is constantly increasing is to have the universe be expanding, in order that over time more and more particles become widely separated and non-interacting. So in a version of the final measurement approach in which the final measurement can select from different cosmologies, it seems plausible that the final measurement may be choosing from a set of histories in which the universe is expanding. Therefore this version of the approach would predict that the final measurement is  likely to select a history where the universe is expanding, even though the universal wavefunction does contain other possible cosmologies.

\cite{Kent} makes a somewhat similar argument, where he notes  that the final measurement approach requires the following conjecture: \emph{`while initially the particles often are localized in the same region and interact, eventually all particles that can decay will have decayed, all particles that are capable of interacting with one another either do interact or become more and more widely separated, non-gravitational interactions become rarer and rarer, and the asymptotic evolution is effectively described by a free quantum field theory'.} This condition ensures that the   limit as $t \rightarrow \infty$ used to define the time of the final measurement is well-defined, since we can perform the measurement at any time after all the particles become completely non-interacting and obtain the same result. Kent also observes that  this sort of infinite expansion appears to be a viable possibility according to our current understanding of cosmology: `\emph{It is supported by some cosmological scenarios that are presently taken seriously, for example (and most cleanly) in “big rip” scenarios.}' \citep{Kent}  However, note that we   have adopted a slightly different dialectic from Kent here. Kent recognises that his model places certain constraints on cosmology but he presents these as assumptions that must be made for the model to work. On the other hand, rather than  assuming that the universe has the right kind of cosmology for Kent's interpretation to work, here we suggest that this behaviour could be derived from an appropriately generalized version of the final-measurement interpretation - in that setting, the universal quantum state may contain branches with different sorts of cosmologies, but the final measurement must select from a set of histories whose cosmologies are compatible with the preservation of generalized records, and therefore there is a good chance in this picture that we will find ourselves in a history which has the property that the universe expands forever so that the ratio  $\frac{N_N}{N_N + N_I}$  is constantly increasing.

\section{The Final-Measurement Account versus the Past Hypothesis \label{ph}}

Let us now compare the approach we have presented here to the standard story in terms of the past hypothesis. First, we note that the arguments made in this paper are only heuristic,   whereas arguments concerning the past hypothesis have been studied to a high level of quantitative detail, and in that sense, the approach to time asymmetry using the past hypothesis  is currently more complete and better supported by quantitative evidence. But a comparison on those grounds is not    very revealing - the past hypothesis is not automatically better just because at this particular moment in time it happens to have been studied more.  Thus in this section we will suppose, for the sake of argument, that all the quantitative details of the suggestions we've made in this article can be worked out to the same level of detail as the past hypothesis approach; and we will contend that if these details do indeed work out, the final-measurement interpretation will in some ways be a better explanation of the arrow of time than the past hypothesis. Of course, assessments of the relative merits of two explanations are always to some degree relative to taste, so we do not necessarily expect that everyone will accept this judgement; however, we do contend that the final-measurement account has a number of theoretical virtues which make it worthy of serious consideration. 

The basic structure of the two explanations seems quite similar: both postulate an underlying time-symmetric dynamics and then add an asymmetry at one end of time or the other. In this sense both are vulnerable to the criticism that they are just putting the asymmetry in by hand (indeed, as Callender's mantra suggests, it is probably impossible to get a temporal asymmetry unless it is in some sense `put in by hand' by including an asymmetry in one's initial postulates). However, there is a crucial difference in the status of the  proposed asymmetry. In the past hypothesis approach, the asymmetry is added by requiring that the initial state of the universe belongs to a special subset of possible initial states. This  has an air of arbitrariness, because the dynamical equations admit many  possible initial conditions and give us no reason to select one rather than another. Indeed such arbitrariness  is inevitable, because the time evolution paradigm on which this approach is predicated insists that dynamics and initial conditions are independent, so there is no hope of ever extracting a reason for one initial condition rather than another from within the dynamical equations defining the theory. In the final-measurement account, on the other hand, the asymmetry is built into the mechanism by which the course of history is selected, and there are no alternative possibilities:  the fact that the final measurement selects from a strongly decoherent set of histories follows directly from the   theory's account of the content of reality, so there is nothing arbitrary or ad hoc here.  Thus the reason for the asymmetry emerges naturally from within the final-measurement interpretation; there is no need to tack it on as an additional hypothesis. 

The two accounts also differ with respect to their unifying power.  The past hypothesis was proposed specifically for the sake of explaining the thermodynamic asymmetry and possibly some of the other temporal asymmetries, insofar as they are derivable from one another. It is therefore doing a substantial chunk of explanatory work, but all of the phenomena which it explains are closely related and can be regarded as offshoots of the problem it was initially postulated to solve. Whereas the final measurement approach was not postulated for any reason to do with thermodynamics or temporal asymmetry: it was intended to solve what Kent refers to as `the Lorentzian classical reality problem' i.e. finding a Lorentz-invariant interpretation of quantum mechanics. Thus if one is willing to accept that it is both a viable solution to the problem of interpreting quantum mechanics  and also an explanation of various temporal asymmetries, it seems to be a more powerful and unifying conjecture than the past hypothesis.  

Moreover, the final-measurement approach seems to do a better job than the past hypothesis with regard to explaining the universality and reliability of thermodynamic behaviour. As argued in particular by \cite{timebomb}, \emph{`if we think that the smooth early universe is just a matter of luck, then we have no reason to expect that the luck will continue, when we encounter new regions of the universe - regions previously too far away to see.}' Price therefore argues that if we are to have any grounds for our expectation that thermodynamical behaviour extends beyond our immediate experience (an expectation which we must have to some extent if we are to allow ourselves to draw any conclusions about reality from our knowledge of the past) we must believe that the special initial state is something more than just a lucky chance. In Price's words, \emph{`The hypothesis is thus being accorded a lawlike status, rather than treated as something that ‘just happens’'.} But if we are working within the time evolution paradigm in which all the laws are dynamical, the initial state can't possibly be lawlike, so it's difficult to see where our confidence in the universality of thermodynamics could come from. By contrast, if the argument of section \ref{thermodynamics} is accepted it follows that the thermodynamical gradient is in fact lawlike, because  the lawlike process which selects the history determining the actual contents of reality is designed in such a way that it is highly likely to select a history which uniformly exhibits entropy-increasing behaviour. So if we believe this account of the selection and actualisation of the course of history, then we have every reason to believe that thermodynamical behaviour is universal, since the requirement for robust records holds everywhere and all times (although as noted earlier, it's possible the records might become less robust close to the end of time). 
 
Now, there have been other  attempts to make the thermodynamical gradient lawlike, such as deriving it from the asymmetric dynamics of spontaneous collapse versions of quantum mechanics \citep{doi:10.1093/bjps/45.2.669}, or  insisting that the past hypothesis is itself nomic \citep{feynman2017character}, or arguing that anti-thermodynamic behaviour cannot ever be remembered \citep{Huckleberry}. But one difficulty with lawlike approaches is that it can be difficult to get the right balance: a nomic approach is liable to end up having the consequence that there is no possibility whatsoever that entropy should ever decrease. And yet we don't typically think it is impossible that anti-thermodynamic behaviour  should occur, and/or be remembered - our best current physics says it is merely unlikely, so we should be suspicious of any approach which rules it out entirely.  The final-measurement account doesn't fall into this trap, because it makes thermodynamical behaviour lawlike without insisting that anti-thermodynamic behaviour is impossible. We have argued that records are far more likely to survive until the final state if they are encoded robustly in the way that follows from entropy-increasing behaviour, so that is the behaviour we should typically expect to see in histories belonging to a strongly decoherent set - but it is still possible that  an event whose records are less robust should appear in a history belonging to a strongly decoherent set, because even if there's only a single system carrying a record of the event, it could be the case that just by good luck it never interacts again. So in this picture it is certainly possible, albeit unlikely, for us to see behaviour which seems anti-thermodynamical, and therefore the final-measurement interpretation explains the great reliability and generality of the second law of thermodynamics without doing violence to our intuition that the second law is valid only statistically.

Finally, we note that the two approaches offer quite different accounts of the underlying reasons for entropy increase.  The final-measurement interpretation is what \cite{timebomb} refers to as the `Causal-General' account of entropy increase, in that it postulates some kind of underlying mechanism which constrains the entropy to increase. (We would not actually describe the final measurement mechanism as `causal', but it has something in common with causal approaches in that it makes the increase of entropy more than just statistical). Whereas the past hypothesis approach is an `Acausal-Particular' account of entropy increase, which attributes entropy increase wholly to a particular fact, i.e. the fact that the initial state of the universe was of some special kind. As argued by Price,  many  Causal-General accounts  do too much: for in addition to the mechanism of entropy increase, they still have to postulate a low entropy initial state in order that the entropy has somewhere to increase from, and after that the causal mechanism seems unnecessary, because standard statistical arguments show that from a low entropy initial state the entropy is very likely to increase in any case. But this  objection does not apply to the final-measurement interpretation, because in this picture we don't have to postulate a low-entropy initial state - we postulate a suitably generic initial quantum state and subsequently select out a set of histories which exhibit entropy-increasing behaviour. Now one might object that we do have to postulate a high-entropy final state instead - after all, if the final quantum state did not have relatively high entropy then there would not be enough information in the outcome of a measurement on it to define the kinds of long complex histories we see in the actual world. But this objection fails because  postulating a high-entropy final state is not, on its own, enough to explain why the entropy increases towards the final state -  conditional on a high-entropy final state, the standard sorts of statistical arguments   tell us that the most likely universe is one where entropy is roughly constant throughout history. So the final measurement approach is not vulnerable to the charge of `doing too much': the mechanism driving entropy increase in the final-measurement picture is doing real work to deliver a gradient rather than a constant high entropy.

\section{Conclusion} 

It is commonly objected that  interpretations of quantum mechanics fail to make `novel predictions'. But if the arguments  in this article are accepted, it would seem that this isn't true in the case of Kent's interpretation  - it does  make a novel prediction, since it correctly predicts a variety of temporal asymmetries for which standard quantum mechanics does not have any explanation. Obviously these asymmetries were known long before Kent's model was formulated, so this  is not the kind of highly coveted novel prediction in which we predict something that we didn't know about prior to the prediction, but Kent's model was not specifically designed to account for temporal asymmetries so we do have an instance of novelty in the   heuristic sense, as articulated by \cite{WORRALL201454}: \emph{`What matters is whether or not the evidence was used in the construction of the theory'.}  One might argue that there is some sense in which the evidence was used in the construction of the theory,  because  Kent's model was designed with decoherence in mind and decoherence is not entirely independent of the other temporal asymmetries, but nonetheless the emergence of an explanation for the other temporal asymmetries out of what was initially intended as a solution to the problem of interpreting quantum mechanics surely  demonstrates novelty of some kind.  So if the suggestions in this paper can be backed up by appropriate quantitative models, they would furnish   a  strong argument in favour of Kent's proposal. 

That said, although we find this account of temporal asymmetry  quite compelling, we are not committed to the claim that it is ultimately the right explanation. After all, much existing research into temporal asymmetry presupposes a time-evolution picture, so it's likely that  if we more actively explore possibilities outside the time-evolution paradigm we will find other  novel approaches to the problem,  and amongst them there  might be an explanation more compelling than the final-measurement approach. Moreover, we recognise that the final-measurement approach  will likely not appeal to those who are already committed to some  other interpretation of quantum mechanics. But the broader point we want to make stands: moving  away from the time evolution paradigm certainly opens up new possibilities for understanding temporal asymmetry, and we would conjecture that it may have new light to shed on other long-standing scientific puzzles as well.

This project also has interesting consequences for the philosophy of scientific explanation.  In the final-measurement approach, rather than explaining features of the universe by appeal to a time evolution law and an initial state, we  explain features of the universe on the basis that events featuring in our observable reality must  create records which persist until the end of time. So we are   giving a teleological explanation - explaining phenomena by appeal to the fact that they are required to create some particular result. Teleological explanations are often regarded as scientifically illegitimate, but given the   underlying time-symmetry of most of the laws of nature, it's unclear why this should be;  if the laws of nature are indeed time-symmetric, there would seem to be no prima facie reason to think that explanations in terms of past conditions are good and explanations in terms of future conditions are bad, nor any cause to suppose that the initial state of the universe should always have explanatory precedence.

Obviously,   we must be careful not to make explanation too easy - one could explain any phenomenon whatsoever by simply saying it is a law that a certain course of events must have the outcome that it in fact has. But this is true of all kinds of explanation: good explanations must do more than simply spit out the desired result, they must have other theoretical virtues like simplicity, comprehensiveness, or unification.  The final-measurement interpretation is a compelling example of a  teleological explanation with significant unifying power: Kent set out to find a coherent interpretation of quantum mechanics, and apparently got out an explanation of a variety of different temporal asymmetries for free. The final-measurement interpretation is therefore a good case study in the rich possibilities that may be uncovered from the exploration of alternative explanatory paradigms.

  \author{Emily Adlam  \thanks{The Rotman Institute of Philosophy, 1151 Richmond Street, London N6A5B7 \texttt{eadlam90@gmail.com} }}

\section{Acknowledgements}

Thanks to Adrian Kent for helpful discussions on this subject. This publication was made possible through the support of the ID \# 62312 grant from the John Templeton Foundation, as part of the project \href{https://www.templeton.org/grant/the-quantum-information-structure-of-spacetime-qiss-second-phase}{‘The Quantum Information Structure of Spacetime’ (QISS)}. The opinions expressed in this publication are those of the author and do not necessarily reflect the views of the John Templeton Foundation.

\end{document}